\documentclass[12pt]{iopart}
\usepackage{iopams}

\begin{document}

\title[The stochastic state selection method]
{The stochastic state selection method combined with
the Lanczos approach to eigenvalues in quantum spin systems}

\author{Tomo Munehisa and Yasuko Munehisa}

\address{Faculty of Engineering, University of Yamanashi, Kofu, Japan, 
400-8511}

\ead{munehisa@yamanashi.ac.jp}

\maketitle

\begin{abstract}
We describe a further development of the stochastic state selection 
method, a new Monte Carlo method we have proposed recently to make
numerical calculations in large quantum spin systems. 
Making recursive use of the stochastic state selection technique 
in the Lanczos approach, we estimate the ground state energy 
of the spin-1/2 quantum Heisenberg 
antiferromagnet on a 48-site triangular lattice. 
Our result for the upper bound of the ground state energy  
is ${-0.1833 \pm 0.0003}$ per bond.
This value, being compatible with values from other work,  
indicates that our method is efficient in calculating energy
eigenvalues of frustrated quantum spin systems on large lattices. 
\end{abstract}


\section{Introduction}
\label{sec1}

In numerical studies of quantum spin systems, one of widely used
approaches is the quantum Monte Carlo method.
This method has helped us greatly to understand
many properties of non-frustrated quantum spin systems, especially of
the spin-1/2 quantum Heisenberg antiferromagnet on bipartite 
lattices~\cite{book1,book2,sand,rev}. 
Nevertheless, the method is ineffective for  
frustrated systems owing to the so-called sign problem. 
It is quite difficult, therefore, 
to draw any definite conclusion from numerical calculations of 
two-dimensional large systems of fermions or frustrated spins.
Yet there are active studies of numerical methods to
investigate these systems.
One of them is the path-integral renormalization group method 
for fermion systems developed by Kashima and Imada~\cite{ki}. Using the Slater
determinant as the basis state they improve the exact diagonalization
method with the truncation of the Hilbert space. 
Another method is proposed by Sorella~\cite{sol}, who extends
the fixed node method making full use of insights into the physics of the 
target system.
One should also note the work of Henelius and Maeshima {\it et al}~\cite{hene, mae} of extending the
density matrix renormalization group method introduced by 
White~\cite{white1, white2}. 

Recently we have developed another Monte Carlo method, 
which we call the {\em stochastic state selection} (SSS) method, 
to calculate eigenvalues in large quantum systems~\cite{mm1,mmss,mm2,mm3}. 
The SSS method has little in common with the ordinary Monte Carlo methods 
because it is not based on importance samplings. A new
type of stochastic algorithm in this method enables us to select   
a relatively small number of elements from a vast vector space 
in a mathematically justified manner. 
Using those selected elements we calculate inner products. 
It is guaranteed that  
we can obtain correct values of these inner products
through the statistically averaging process. 

So far we have used the SSS method in combination with the power method,
since it is a simple and straightforward way
of applying the SSS method to the numerical study on energy eigenvalues.
In this paper we employ the Lanczos approach instead of the power method.
The Lanczos method gives us reliable results on small lattices for which 
we can keep every state in the vector space; for larger lattices
we need to make some truncations because the vector space becomes huge.
Use of truncated states in the Lanczos method, however, has been
unsuccessful in either theoretically justifying its methodology or 
in obtaining better numerical eigenvalues~\cite{riera}. 
Our purpose in this work is to show that for a large frustrated system  
we can evaluate the coefficients necessary in the Lanczos approach
by means of the SSS method.

As a concrete example, we calculate an energy eigenvalue 
on a $N_s$-site triangular lattice for the ground
state energy of the spin-1/2 quantum Heisenberg antiferromagnet. 
The Hamiltonian of the system is 
\begin{equation}
\hat H = \frac{J}{4} \sum_{(i,j)} \bsigma_i \ \bsigma_j \,,
\label{hamil}
\end{equation}
where $\bsigma_i$ denotes Pauli spin matrices on the $i$-th site of  
a triangular lattice with $N_s$ sites  
and the sum runs over all $N_b(=3N_s)$ bonds of the lattice. 
The coupling $J$ is set to 1 throughout this paper.
The reason why we study this system here is that, as is well-known,
it is one of the systems that is strongly frustrated in two 
dimensions~\cite{rev,bernu,oguchi,nishim,leung,sindz,capri,farnell,singh}. 
As far as we know, 
reported results on the model by means of the exact diagonalization
method are only for lattices smaller than or equal to 36 sites~\cite{bernu}. 
The largest lattice size is also 36 in our previous work~\cite{mm2}, where 
the recursive SSS method and the power method are employed.
In the present study with an improved approach, we first calculate
the lowest energy eigenvalue of the $N_s=27$, $S_z=1/2$ system 
in order to confirm that our new approach works well. 
For this system we mimic conditions of a 
virtual small computer so that we can conclude the results obtained in
reduced vector space are reliable (table \ref{t0}). 
We then proceed to the $N_s=48$, $S_z=0$ system (table \ref{t1}), 
for which our result is $E/JN_b = -0.1833 \pm 0.0003$. 
A happy combination of the recursive SSS method with the Lanczos
approach in this paper 
enables us to estimate the ground state energy of the 48-site system. 

The plan of this paper is as follows. In the next section we explain our
method. Section \ref{sec3} is devoted to a detailed account of the trial
state on triangular lattices. 
Calculations with stochastic state selections are described 
in section \ref{sec4}. 
The final section gives a summary and discussions.  
  
\section{Method}
\label{sec2}

In this section we present brief descriptions of the SSS method and our
Lanczos approach.

\subsection{SSS method}

The stochastic state selection is realized by a number of random variables.
Let us expand a state ${\mid \phi \rangle}$ by some basis 
${\{ \mid i \rangle \}}$, 
${\mid \phi \rangle} = {\sum_{i=1}^{N} c_i \mid i \rangle}$.
Then we generate a random variable $\eta_i$ following to the
{\em on-off probability function}
\begin{equation}
P_i(\eta) \equiv 
\frac{1}{a_i}\delta(\eta -a_i) +(1- \frac{1}{a_i})\delta(\eta) 
\qquad \frac{1}{a_i} \equiv \min \left( 1, \frac{|c_i|}{\epsilon} \right).
\end{equation}
A positive parameter $\epsilon $ common to all ${P_i(\eta)}$ 
$(i=1,2,\cdots, N)$ controls the reduction rate. 
Note that $\eta_i=a_i$ or $\eta_i=0$ and
statistical averages are ${\langle \! \langle \eta_i \rangle \! \rangle} = 1$
and ${\langle \! \langle \eta_i^2 \rangle \! \rangle} = a_i$.
A {\em random choice operator} $\hat{M}_{\{\eta\}} $ is defined by  
\begin{equation}
\hat{M}_{\{\eta\}} \equiv
\sum_{j=1}^N \mid j \rangle \eta_j \langle j \mid \,.
\end{equation}
Using this $\hat{M}_{\{\eta\}} $ 
we obtain a state $\hat{M}_{\{\eta\}} {\mid \phi \rangle }= 
{\sum c_i \eta_i \mid i \rangle}$, which has fewer non-zero elements 
than ${\mid \phi \rangle}$.
An expectation value ${\langle \phi \mid \hat O \mid \phi \rangle }$
with an operator $\hat O$ is exactly equal to the statistical average
${\langle \! \langle \ \langle \phi \mid \hat O \hat{M}_{\{\eta\}} 
\mid \phi \rangle \ \rangle \! \rangle }$. 
When we want to emphasize that different random choice operators are used,
we will denote them by $\hat{M}_{\{ \eta^{(k)} \}}$ $(k=1,2, \cdots)$. 

\subsection{Lanczos approach}

In the Lanczos approach we start from a state $ {\mid \psi^{(1)}\rangle} $ 
and calculate, for some $L$, orthogonal states 
$ {\mid \psi^{(m)}\rangle} \ (m=2,3, \cdots, L)$ together with 
$\alpha_m \ (m=1,2, \cdots, L)$ and $\beta_m \ (m=1,2,\cdots, {L-1})$, 
\begin{equation}
 \mid \psi^{(m)} \rangle \equiv \frac{1}{\beta_{m-1}} 
\{
\hat Q \left( \alpha_{m-1}\right) \mid \psi^{(m-1)}\rangle -\beta_{m-2}
\mid \psi^{(m-2)}\rangle \} \,,
\end{equation}
\begin{equation}
\alpha_m \equiv \langle \psi^{(m)} \mid \hat{Q} \left( 0 \right)\mid 
\psi^{(m)}\rangle \,, 
\label{am}
\end{equation}
\begin{equation} 
\beta_m \equiv \sqrt{ \langle \psi^{(m)} \mid  
 \hat{Q} \left(\alpha_m \right) ^2 
\mid \psi^{(m)}\rangle -\beta_{m-1}^2} \,, 
\label{bm} 
\end{equation}
where we define $\beta_0 \equiv 0$ and $ \hat{Q} \left(\alpha \right)
\equiv \hat H - \alpha$ with the Hamiltonian $\hat H$, 
so that we obtain the tridiagonal matrix 
\begin{equation}
\cal{A}_{\it L} \equiv
\left[ \begin{array}{ccccccc}
\alpha_1 & \beta_1 & 0 & 0 & \cdots &0 & 0 \\
\beta_1 & \alpha_2 & \beta_2  & 0 & \cdots &0 &  0 \\
0 & \beta_2 & \alpha_3 & \ddots & \cdots & 0 & 0 \\
\vdots &  0 & \ddots &   \ddots & \ddots & \vdots & \vdots \\
0 & 0 & \vdots  & \ddots & \ddots & \ddots &  0 \\
0 & 0 & 0 & \cdots & \ddots & \ddots & \beta_{L-1} \\
0 & 0 & 0 & \cdots & 0 & \beta_{L-1}& \alpha_L
\end{array}\right] .
\label{cala}
\end{equation}
The $L$-th approximate eigenvalue for the ground state is given by
the lowest eigenvalue of $\cal{A}_{\it L}$. 
Let us denote the lowest eigenvalue of $\cal{A}_{\it L}$ by 
$\tilde{\alpha}_L$ and its eigenvector by $\bi{u}^{(L)}\equiv \left[
u^{(L)}_1, \  u^{(L)}_2, \ \cdots, \ u^{(L)}_L \right] ^{\rm{T}}$
hereafter.
Once we know $\cal{A}_{\it L}$, 
$\tilde{\alpha}_L$ and $\bi{u}^{(L)}$ for some $L$, 
we can in principle evaluate $\beta_L$ and $\alpha_{L+1}$ from relations
\begin{equation}
\langle \tilde{\psi}^{(L)} \mid 
\hat{Q} \left(\tilde{\alpha}_L \right) ^2
\mid \tilde{\psi}^{(L)} \rangle = \{ u^{(L)}_L \}^2 \beta_L ^2 \,, 
\label{clbet}
\end{equation}
\begin{equation} 
\langle \tilde{\psi}^{(L)} \mid 
\hat{Q} \left(\tilde{\alpha}_L \right)\hat{Q} \left( 0 \right)\hat{Q} 
\left(\tilde{\alpha}_L \right)
\mid \tilde{\psi}^{(L)}\rangle = \{u^{(L)}_L\}^2 \beta_L ^2 \alpha_{L+1} \,, 
\label{clalp}
\end{equation}
with a state 
\begin{equation}
\mid \tilde{\psi}^{(L)}\rangle \equiv
\sum_{m=1}^{L} u_m^{(L)}\mid \psi^{(m)}\rangle \,.
\end{equation} 
This would lead us to a larger matrix $\cal{A}_{\it L+1}$ 
and its lowest eigenvalue $\tilde{\alpha}_{L+1}$ would  
give us a better estimate of the ground state energy.

In order to perform a successful numerical evaluation with a small value
of $L$, 
it is necessary to make ${\mid \psi ^{(1)} \rangle}$ as good as possible.
Remember that it is difficult to directly calculate $\alpha_m$ and 
$\beta_m$ for large systems with such a ${\mid \psi ^{(1)} \rangle}$,
because our available computer memory resources would not be enough to
keep whole elements of ${\hat H \mid \psi^{(1)}\rangle}$. 
We therefore truncate each state by operating
an $\hat{M}_{\{ \eta \}}$ to it. Details of our stochastic 
selection will be mentioned in section \ref{sec4}.

\section{Trial state}
\label{sec3}

Before describing the stochastic selection in our calculation, it would
be necessary to explain how we prepare the trial state 
$\mid \psi^{(1)} \rangle$ to start with.
In this section and in the next section we concentrate our attention on
the $N_s=48$ case, which involves more technical issues than the
$N_s=27$ case does. 
 
First let us comment on our basis by which we describe the states 
${\mid \psi^{(m)} \rangle}$.
The representation we use is 
the conventional one where a state is represented by $z$-components 
of all spins of the system. 
Here we add an assumption on symmetries. 
On a triangular lattice with 48 spins there are 48
translation operators as well as 6 rotation and 2 inversion ones
that commute with the Hamiltonian (\ref{hamil}). 
Since we expect that the ground state is invariant under these
operations~\cite{bernu}, 
we construct a basis ${\{ \mid \Phi_i \rangle \}}$ which ensures 
all of these invariances. Note that each basis state $ \mid \Phi_i \rangle $
therefore contains maximally 576 ($= {48 \times 6 \times 2} $) 
degenerate spin configurations in it. Total number of $\mid \Phi_i
\rangle$ with $S_z = 0$ therefore is about $6 \times 10^{10}$. 

Now we come to the starting point of our numerical work, which is to
calculate coefficients of ${\mid \psi^{(1)} \rangle}$
with the basis stated above. We do it  
in the same way that we did in our previous work~\cite{mm2}, 
where we obtained an approximate ground state ${\mid \psi_{\rm A} \rangle} $ 
for the spin system on a 36-site triangular lattice. 
The only difference is that we include as many degenerate Ising-like
configurations as possible in the initial trial state this time. 
The basic idea for this improvement comes from the Wannier's rigorous 
proof\cite{wan} 
that a classical antiferromagnetic Ising system on a triangular 
lattice is heavily degenerated in its minimum energy, namely its energy 
at zero temperature, which is $-N_s/4$ for the $N_s$-site system.
 
Using the conventional Monte Carlo method at low temperature ($T=0.5$), 
where the classical energy is used as the Boltzmann weight,
we pick up states with the classical minimum energy, which is
$-12$ for the $N_s=$48-site system. 
We find ${13\,087}$ ${\mid \Phi_i \rangle}$'s that fulfill conditions 
${\langle \Phi_i \mid \hat{H} \mid \Phi_i \rangle} = -12.0$ and $S_z=0$. 
We then form an initial trial state ${\mid \Psi_0 \rangle}$ by linearly 
combining all of them with the equal weight ${1/\sqrt{13087}}$. 
Within this partial Hilbert space with ${13\,087}$ basis states, we next
pursue the state ${\mid \Psi_{\rm t} \rangle}$ which has the lowest energy in
the conventional exact diagonalization. 
We observe that ${\langle \Psi_{\rm t} \mid
\hat H \mid \Psi_{\rm t} \rangle} = -20.1$.

The final stage to calculate a trial state ${\mid \psi^{(1)} \rangle}$
is to repeat following procedures until the obtained value does not
change within five decimal digits.
\begin{enumerate}
\item Extend the partial Hilbert space ${\{\mid \Phi_i \rangle \ ; \ 
\langle \Phi_i \mid \hat H \mid \Phi_i \rangle = -12.0,\  S_z = 0\}}$
by operating $\hat H$ a few times
      to ${\mid \Psi_{\rm t} \rangle}$ until the available computer
      memory is exhausted. The maximum number of basis states we can permit
      is $\sim 1 \times 10^8$.
\item Within the Hilbert space determined in (i), pursue the state 
      ${\mid \Psi'_{\rm t} \rangle}$ with which 
      ${\langle  \Psi'_{\rm t} \mid \hat H \mid \Psi'_{\rm t} \rangle }$
      is as low as possible.
\item Form a state ${\mid \Psi''_{\rm t} \rangle}$
      by dropping small coefficients of ${\mid \Psi'_{\rm t} \rangle}$.
      We usually request that the size of the reduced Hilbert 
      space, which is spanned by basis states included in 
      ${\mid \Psi''_{\rm t} \rangle}$, should be 
      a few percents of the one obtained in (i).
\item Replace ${\mid \Psi_{\rm t} \rangle}$ by ${\mid \Psi''_{\rm t} \rangle}$.
\end{enumerate} 
After this calculation we obtain the trial state 
${\mid \psi^{(1)} \rangle}$ from the last ${\mid \Psi'_{\rm t}
\rangle}$. 
We observe that 
${\mid \psi^{(1)} \rangle} $ comprises 75\,746\,657 basis states.
Then we obtain $\alpha_1 = {\langle \psi^{(1)} \mid \hat H 
\mid \psi^{(1)} \rangle} = -25.950$. It should be noted that we can calculate 
this inner product exactly because we do not need to keep the `outer'
part of ${\hat H \mid \psi^{(1)}\rangle} $ which are orthogonal to    
the vector space attached to ${\mid \psi^{(1)}\rangle}$.

\section{Calculations}
\label{sec4}

With the Hamiltonian (\ref{hamil}) and the trial state 
${\mid \psi^{(1)}\rangle}$ described in the
previous section, we calculate the matrix elements in ${\cal{A}}_L$
up to $L=4$. This section is to show in detail 
how we carry out the calculations with stochastic selections.
Note that all the values and the conditions stated below in this section
are those for the $N_s=48$ lattice.
Comments on the results summarized in tables~\ref{t0} and \ref{t1}
will be presented in the next section.
 
First step here is to estimate $\beta_1$ and $\alpha_2$ so that we
can solve the eigenvalue problem with ${\cal{A}}_2$.
Using the recursive SSS~\cite{mm2}, 
we calculate the most probable value of $\beta_1^2  = 
{\langle \psi^{(1)} \mid \hat{Q} \left(\alpha_1 \right)^2
\mid \psi^{(1)} \rangle} $ from the statistical average of 
\begin{equation}
\langle \psi^{(1)} \mid \cdot \  \hat{Q}\left(\alpha_1 \right)
 \hat{M}_{\{\eta ^{(2)}\}} \hat{Q}\left(\alpha_1 \right)
\hat{M}_{\{\eta ^{(1)}\}}\mid \psi^{(1)}\rangle \,.
\label{b1b1} 
\end{equation} 
Here we insert a symbol $\cdot$ after
${\langle \psi^{(1)}  \mid}$ in order to represent that we
calculate the inner product between ${\mid \psi^{(1)} \rangle}$ and 
${\hat{Q}\left(\alpha_1 \right) \hat{M}_{\{\eta ^{(2)}\}}
\hat{Q}\left(\alpha_1 \right) 
\hat{M}_{\{\eta ^{(1)}\}}\mid \psi^{(1)}\rangle }$.
Note that each random choice operator in the recursive SSS method depends
on the preceding intermediate state.
We generate each $\hat{M}_{\{\eta \}}$
adjusting the value of $\epsilon$ to be as small as possible 
for a Pentium IV machine equipped with a 2 Giga byte memory.  
Our result from 22 samples 
is $\beta_1 = 1.7082 \pm 0.0021$,
where the error is estimated by the standard deviation of the data.  
When we evaluate $\alpha_2$, we approximate ${\mid {\psi} ^{(2)} \rangle} 
= {\hat Q \left( \alpha_1 \right) \mid \psi^{(1)} \rangle / \beta_1}$ by 
${\hat Q \left( \alpha_1 \right) \hat{M}_{\{\eta\}}\mid{\psi} ^{(1)} 
\rangle / \beta_1}$. Namely we calculate 
\begin{equation}
\langle \psi^{(1)} \mid \cdot \ \hat Q \left( \alpha_1 \right) 
\hat{M}_{\{\eta ^{(3)}\}} \hat Q \left( 0 \right)  \hat{M}_{\{\eta ^{(2)}\}}
\hat Q \left( \alpha_1 \right) \hat{M}_{\{\eta ^{(1)}\}}
\mid \psi^{(1)} \rangle
\label{b1b1a2}  
\end{equation}
to obtain ${ \beta_1^2 \alpha_2}$. 
The statistical average from 40 samples gives
us $\alpha_2 = {-12.007 \pm 0.066}$. 
Here we take account of both the statistical
error for (\ref{b1b1a2}) and the error from $\beta_1$. 
It is easy to find that $\tilde{\alpha}_2 = -26.1559 $, 
$u^{(2)}_1=0.99279$ and $u^{(2)}_2=-0.11986$   
with the above values of $\alpha_1$, $\beta_1$ and $\alpha_2$.
Error estimations on these quantities are carried out as follows. 
Let $f$ be $\tilde{\alpha}_2$, $u^{(2)}_1$ or $u^{(2)}_2$. 
We assume, with errors ${\Delta \beta_1}$ and ${\Delta \alpha_2}$,  
\begin{equation}
\fl
f(\alpha_1, \beta_1 \pm  \Delta \beta_1, \alpha_2 \pm \Delta \alpha_2)
\simeq f(\alpha_1,\beta_1,\alpha_2 ) \pm \sqrt{ 
\left[ \frac{\partial f}{\partial \beta_1} \Delta \beta_1 \right]^2+
\left[ \frac{\partial f}{\partial \alpha_2} \Delta \alpha_2 \right]^2} 
\end{equation} 
and numerically calculate 
$\displaystyle 
\left[ \frac{\partial f}{\partial \beta_1}\Delta \beta_1 \right]$ and 
$\displaystyle 
\left[ \frac{\partial f}{\partial \alpha_2} \Delta \alpha_2 \right]$
by solving eigen problems of matrices 
\begin{equation}
\left[ \begin{array}{cc}
\alpha_1 & \beta_1 \pm \Delta \beta_1 \\
\beta_1 \pm  \Delta \beta_1 & \alpha_2 \end{array}\right] \,, \ \ \  
\left[ \begin{array}{cc}
\alpha_1 & \beta_1\\
\beta_1 &\alpha_2 \pm \Delta \alpha_2 \end{array}\right] \,. 
\end{equation}
 
Then we proceed to estimate $\beta_2$ and $\alpha_3$ using (\ref{clbet})
and (\ref{clalp}) with $L=2$. 
We approximate ${\mid \tilde{\psi} ^{(2)} \rangle} = {u_1^{(2)} \mid \psi_1
\rangle + u_2^{(2)}\mid \psi_2 \rangle} $ by
${\hat P_2 \left(\hat{M}_{\{\eta\}}\right)\mid \psi^{(1)} \rangle}$, where 
\begin{equation} 
\hat P_2 \left(\hat{M}_{\{\eta\}}\right) \equiv  
u_1^{(2)}+\frac{u_2^{(2)}}{\beta_1}
\hat Q (\alpha_1)  \hat{M}_{\{\eta \}}  \,. 
\end{equation}
We calculate statistical averages of  
\begin{equation}
\fl
\langle \psi^{(1)} \mid \cdot \  
\hat P_2 \left( \hat{M}_{\{\eta ^{(7)}\}} \right)
\hat Q \left( \tilde{\alpha}_2 \right) \hat{M}_{\{\eta ^{(6)}\}}
\hat Q \left( \tilde{\alpha}_2 \right) \hat{M}_{\{\eta ^{(5)}\}} 
\hat P_2 \left( \hat{M}_{\{\eta ^{(4)}\}} \right)
\mid \psi^{(1)} \rangle 
\end{equation}
and 
\begin{equation}
\fl
\langle \psi^{(1)} \mid \cdot \  
\hat P_2 \left(\hat{M}_{\{\eta ^{(9)}\}} \right)
\hat Q \left( \tilde{\alpha}_2 \right) \hat{M}_{\{\eta ^{(8)}\}}
\hat Q \left( 0 \right) \hat{M}_{\{\eta ^{(6)}\}}
\hat Q \left( \tilde{\alpha}_2 \right) \hat{M}_{\{\eta ^{(5)}\}} 
\hat P_2 \left(\hat{M}_{\{\eta ^{(4)}\}} \right)
\mid \psi^{(1)} \rangle 
\end{equation}
to evaluate ${\{ u^{(2)}_2 \}^2 \beta_2 ^2}$ and ${\{ u^{(2)}_2 \}^2
\beta_2 ^2 \alpha_3}$, respectively. 
From 34 and 128 samples of these quantities, we estimate $\beta_2$ and
$\alpha_3$. We also evaluate their errors, taking the error from $u^{(2)}_2$ 
into account.   
Calculations for $\tilde{\alpha}_3$ and $u^{(3)}_m$'s are then straightforward
including error estimations. 
For the results, see table~\ref{t1}.

In estimations of $\beta_3$ and $\alpha_4$ we calculate inner 
products between
\begin{equation}
\hat Q \left( \tilde{\alpha}_2 \right) \hat{M}_{\{\eta ^{(12)}\}} 
\hat P_3 \left( \hat{M}_{\{\eta ^{(11)}\}} , \hat{M}_{\{\eta ^{(10)}\}}\right)
\mid \psi^{(1)} \rangle 
\end{equation}
and 
\begin{equation}
\hat R \left( \hat{M}_{\{\eta ^{(16)}\}} \right)
\hat Q \left( \tilde{\alpha}_2 \right)  \hat{M}_{\{\eta ^{(15)}\}} 
\hat P_3 \left( \hat{M}_{\{\eta ^{(14)}\}} , \hat{M}_{\{\eta ^{(13)}\}}\right)
\mid \psi^{(1)} \rangle \,, 
\end{equation}
where we use abbreviations  
\begin{equation}
\fl
\hat P_3 \left( \hat{M}_{\{\eta\}},  \hat{M}_{\{\eta ' \}} \right) \equiv  
u_1^{(3)}-\frac{\beta_1}{\beta_2}u_3^{(3)}
+\left\{ \frac{u_2^{(3)}}{\beta_1}
+\frac{u_3^{(3)}}{\beta_1 \beta_2} \hat Q \left( \alpha_2 \right) 
\hat{M}_{\{\eta \}}\right\} \hat Q \left( \alpha_1 \right) 
\hat{M}_{\{\eta ' \}}  
\,, 
\end{equation}
\begin{equation}
\fl
\hat R \left( \hat{M}_{\{\eta \}} \right) \equiv 
\left\{
\begin{array}{ll} 1 & \mbox{to calculate $\{u_3^{(3)}\}^2 \beta_3^2 $} \\
\hat Q\left( 0 \right) \hat{M}_{\{\eta \}} &\mbox{to calculate
 $\{u_3^{(3)}\}^2 \beta_3^2 \alpha_4$} \end{array} \right.
\,,
\end{equation}
for convenience,
so that we can generate as many samples as possible within a limited CPU time.
The results for $\beta_3$ and $\alpha_4$, which are obtained from 6240 and
4069 samples respectively, are in table~\ref{t1}.
The maximum number of the selected basis states amounts to 
$\sim~1.1~\times~10^8$.
\footnote
{Similar calculations for $\beta_4$ and $\alpha_5$ 
would be possible with a swifter computer. 
We estimate that the CPU time necessary to calculate
these quantities with our Pentium IV machine is about 300 times 
as long as the CPU time we spent for $\beta_3$ and $\alpha_4$.}
    
Finally, one comment would be necessary from technical point of view. 
It should be noted that there are variety of ways to calculate samples
using the recursive SSS method for the following two reasons.
\begin{enumerate}
\item    
Suppose we calculate $\beta_1^2$, for example.
For this purpose we calculated inner products (\ref{b1b1}).   
Yet, the statistical average of inner products between 
${\hat Q \left(\alpha_1 \right) 
\hat{M}_{\{\eta^{(1)}\}} \mid \psi^{(1)}\rangle}$ and 
${\hat Q \left( \alpha_1 \right) 
\hat{M}_{\{\eta^{(17)}\}}\mid \psi^{(1)}\rangle}$
will also give us the same quantity.
\item
Locations of random choice operators are not uniquely determined. 
For instance, it is possible to employ 
\begin{equation} 
\hat P'_2 \left(\hat{M}_{\{\eta\}}\right) \equiv  
\left[u_1^{(2)}+\frac{u_2^{(2)}}{\beta_1}
\hat Q (\alpha_1) \right] \hat{M}_{\{\eta \}}  \,, 
\end{equation}
and 
\begin{equation} 
\fl
\hat P'_3 \left( \hat{M}_{\{\eta\}}, \hat{M}_{\{\eta ' \}} \right) \equiv  
\left[u_1^{(3)}-\frac{\beta_1}{\beta_2}u_3^{(3)}
+\left\{ \frac{u_2^{(3)}}{\beta_1}
+\frac{u_3^{(3)}}{\beta_1 \beta_2} \hat Q \left( \alpha_2 \right) 
\right\} \hat{M}_{\{\eta \}} \hat Q \left( \alpha_1 \right)\right] 
\hat{M}_{\{\eta ' \}}  
\,, 
\end{equation}
instead of ${\hat P_2 \left( \hat{M}_{\{\eta\}}\right)}$ and 
${\hat P_3 \left( \hat{M}_{\{\eta\}}, \hat{M}_{\{\eta ' \}} \right)}$, 
respectively.
\end{enumerate}
Although statistical averages obtained in these ways will theoretically
agree with each others, their standard deviations might be different.
It is not {\it a priori} clear what way is best for numerical calculations.  
In order to present an example which shows how much difference is 
actually observed, we calculate 300 samples of 
\begin{equation}
\fl
\langle \psi^{(1)} \mid \cdot 
P'_3 \left( \hat{M}_{\{\eta ^{(23)}\}} , \hat{M}_{\{\eta ^{(22)}\}}\right)
\hat Q \left( \tilde{\alpha}_3 \right)  \hat{M}_{\{\eta ^{(21)}\}} 
\hat Q \left( \tilde{\alpha}_3 \right)  \hat{M}_{\{\eta ^{(20)}\}}  
P'_3 \left( \hat{M}_{\{\eta ^{(19)}\}} , \hat{M}_{\{\eta ^{(18)}\}}\right)
\mid \psi^{(1)} \rangle \,, 
\end{equation}
whose statistical average also should give ${\{ u^{(3)}_3 \}^2 \beta_3 ^2} $.
The result is $\beta_3 = {10.8 \pm 11.0}$ with a fixed value of  
$\epsilon= 0.0025$, and we observe almost the same variance as 
in the measurement with  
$\hat P_3 \left( \hat{M}_{\{\eta\}},  \hat{M}_{\{\eta ' \}} \right)$.

\section{Summary and discussions}
\label{sec5}
 
In this paper we combine our recursive SSS method with the
Lanczos approach so that we can estimate the ground state energy of 
the spin-1/2 quantum Heisenberg antiferromagnet on a
48-site triangular lattice.

In order to examine whether the method works well, we study the 27-site 
system for which the ground state energy is exactly known to be 
$E/JN_b = -0.1867404$. 
Our results for the 27-site system in table~\ref{t0} give a satisfying 
upper limit $E/JN_b \leq {\tilde{\alpha}_4/JN_b} = {-0.18506 \pm 0.00037}$. 
It should be noted that we obtain this upper limit within a limited partial
Hilbert space. 

Now we summarize our results for the 48-site system.  
Starting from a state ${\mid \psi^{(1)} \rangle} $ with ${\sim 7.6 \times
10^7}$ basis states, each of which being a representative of maximally 576 
translation-, rotation- and inversion-invariant configurations,
we successfully calculate 
elements of the tridiagonal matrix ${\cal{A}}_4$, namely $\alpha_1$,
$\alpha_2$, $\alpha_3$, $\alpha_4$, $\beta_1$, $\beta_2$ and $\beta_3$
presented in table~\ref{t1}.  
Our best estimate for the upper bound of the ground state energy is
given by $\tilde{\alpha}_4$. The result per bond is 
${\tilde{\alpha}_4/JN_b} ={-0.1833 \pm 0.0003}$.

This value should be compared with values obtained by other methods. 
In a figure presented by Capriotti {\it et. al.}~\cite{capri}, who
study the system using the Green function Monte Carlo method, we see
that ${E/JN_b} \simeq -0.185$ when $N_s=48$.
The variational Monte Carlo study~\cite{singh} presents a value
${E/JN_b} = {-0.185 \pm 0.001}$ for the 48-site system. 
Richter {\it et. al.}~\cite{rev}, on the other hand, made the finite-size
extrapolation using the results for $N_s=24$, 30 and 36 obtained by the
exact diagonalization method. 
Using the scaling formula 
${e_0 \left(\sqrt{N_s}\right)}=A_0 + {A_3/\left(\sqrt{N_s}\right)^3} 
+ {{\cal{O}}\left(N_s^{-2}\right)}$ for the lowest energy per site 
$e_0 \equiv {E/JN_s} = 3 {E/JN_b}$,  
they obtain ${A_0/3}={e_0(\infty)/3}=-0.1842$.\footnote{
Simple extrapolation using our results on $N_s=36$ and 48 lattices 
yields an upper bound ${E/JN_b \leq -0.1790 \pm 0.0025}$ 
in the ${N_s \rightarrow \infty}$ limit. }
The value for the 48-site system in this formula, which we calculate from 
this $A_0$ and the value ${e_0(\sqrt{36})/3}=-0.1867912$~\cite{bernu}, 
is ${e_0(\sqrt{48})/3 = -0.1859}$.
Thus we see that our result is consistent with those obtained by the
Green function Monte Carlo method and the variational Monte Carlo method.
This indicates that the recursive SSS method combined with the Lanczos
approach proves to be one of techniques applicable to frustrated
quantum systems.

A few remarks are in order.

Applications of the SSS method in study of other models, 
especially of frustrated ones, are in prospect. 
For example, we have studied the 64-site Shastry-Sutherland model 
combining the simple SSS method with the power method\cite{mmss}, 
where we obtained results to indicate that the intermediate phase exists.
Much improved results on this model are expected 
with our new method presented in this paper.

The result ${\tilde{\alpha}_4/JN_b} ={-0.1833 \pm 0.0003}$ 
on the 48-site triangular lattice is much more precise compared with our
previous result $\displaystyle {E_{\rm fit}/JN_b} =
-0.1856 \pm \begin{array}{l}
{\mbox{\small 0.0009}} \\ {\mbox{\small 0.0006}} \end{array}$
on the 36-site lattice\cite{mm2}.
One reason for this improvement is that, instead of the power method, 
we adopt a Lanczos approach
with which the fitting procedure is not necessary.
Another reason is that we improve the trial state
noticing the existence of heavily degenerate Ising-like states 
that have the lowest
energy in the classical Ising system on the triangular lattice.
This idea, inspired by the Wannier's proof\cite{wan} on the classical
Ising system, is encouraged by observations in numerical study of quantum
systems on small lattices. For $N_s =12,$ 21 and 27 lattices we observe 
that a large part of the ground state is 
formed by those degenerate Ising-like basis states. 
 
The final remark is on the merits of the SSS method. 
First of all, we emphasize that this method
is mathematically justified. It is guaranteed that expectation values
for any operator and any state are given correctly by the statistical 
averages.
Secondly our method is quite general in a sense that the SSS method can
be substituted for the importance samplings in usual Monte Carlo
methods~\cite{book1,book2}. This should be compared with the variational
techniques, for which one needs deep insights on properties
of the system under study\cite{sindz,capri}.
%
Thirdly one can easily join the SSS technique to various  
methods such as the power method, the Lanczos method and the variational
method. We expect this technique will be helpful to improve many
methods used in numerical studies.

\begin{table}[h]
\caption{Estimated values of $\alpha$s and $\beta$s in the Lanczos
approach for the $N_s=27$ system, to which we know that 
$E_{\rm exact} = -15.125972$\cite{bernu}. 
For this lattice we simply collect all $S_z=1/2$ states to form a basis.
The size of the full Hilbert space is therefore
 $_{27}$C$_{13}$=20\,058\,300.
We start from a trial state with 1\,072\,935 basis states obtained after the
 procedures in section \ref{sec3}, for which $\alpha_1 =  -14.5680$. 
The last three rows in the table present exact values which are
 calculated in the full Hilbert space without any stochastic
 selections.
When we calculate $\alpha$s and $\beta$s 
inside a limited partial Hilbert space,
we set the maximum size of the space to be 4\,500\,000. 
%
%
}
\label{t0}
\begin{center}
\begin{tabular}{cccc} \hline
$L$ & 2 & 3 & 4 \\ \hline \hline
$\beta_{L-1}$ & $1.5405$ $\pm$  $0.0015$ & $6.14$  $\pm$  $0.18$ &  
$6.09$  $\pm$  $0.55$\\ \hline
$\alpha_L$ & $-4.999$  $\pm$  $0.014$ & $-3.53$  $\pm$  $0.39$ & $-1.8$
$\pm$  $1.5$\\ \hline
$\tilde{\alpha}_L$ & $-14.8099$ $\pm $ $0.0023$ & $-14.927$ $\pm $  $0.010$
& $-14.990$  $\pm $   $0.030$\\ \hline \hline 
%
%
$\beta_{L-1}^{{\rm exact}}$ 
& $1.54010$ & $6.23049$   &  $6.06325$ \\ \hline
$\alpha_L^{{\rm exact}}$ 
& $-5.01580$ & $-3.72843$  & $-1.85916$ \\ \hline
$\tilde{\alpha}_L^{{\rm exact}}$ 
& $-14.8101$ & $-14.9354$  & $-15.0061$ \\ \hline
\end{tabular}
\end{center}

\caption{Estimated values of $\alpha$s and $\beta$s in the Lanczos
 approach for the $N_s=48$ system. The maximum number of basis states is
 limited to $1.1 \times 10^8$. 
As is described in section \ref{sec3} we 
first calculate a trial state with 75\,746\,657 basis states, for which 
$\alpha_1$ is $-25.950$. Then, following   
equations (\ref{b1b1}) and (\ref{b1b1a2}) in section \ref{sec4}, 
we estimate $\beta_1$ and $\alpha_2$. Thirdly we calculate 
$\tilde{\alpha}_2$ and $u_1^{(2)}, u_2^{(2)}$ by an exact diagonalization.
In a similar way to that stated in section \ref{sec4} in detail, 
we evaluate quantities for $L=3$ and $L=4$.} 

\label{t1}
\begin{center}
\begin{tabular}{cccc} \hline
$L$ & 2 & 3 & 4 \\ \hline \hline
$\beta_{L-1}$ & $1.7082$ $\pm$  $ 0.0021$ & $12.63$  $\pm$  $ 0.16$ &  
$13.2$  $\pm$  $2.0$\\ \hline
$\alpha_L$ & $-12.007$  $\pm$  $0.066$ & $-1.43$  $\pm$  $0.45$ & $12.1$
$\pm$  $ 6.6$\\ \hline
$\tilde{\alpha}_L$ & $-26.1559$ $\pm $ $0.0011 $ & $-26.3189$ $\pm $  $0.0096 $
& $-26.393$  $\pm $   $0.049 $\\ \hline
$u_1^{(L)}$& $+0.99279$ $\pm $ $0.00007 $  & $+0.9719$  $\pm $   $0.0016 $
& $+0.955$ $\pm $  $0.013 $ \\ \hline
$u_2^{(L)}$&  $-0.11986$ $\pm $ $0.00056 $ & $-0.2101$  $\pm$   $0.0051 $
& $-0.248$  $\pm $   $0.024 $\\ \hline
$u_3^{(L)}$& ---&
$+0.1066$ $\pm $ $0.0048 $ & $+0.153$ $\pm $   $0.030 $ \\ \hline
$u_4^{(L)}$ &--- & --- & $-0.052$ $\pm $   $0.024 $ \\ \hline
\end{tabular}

\end{center}
\end{table}

\end{document}